\documentclass[prl,twocolumn,aps,showpacs]{revtex4}
\begin{document}

\title{Non-quantum Entanglement Resolves a Fundamental Issue in 
 Polarization Optics}
\author{B. Neethi Simon}
%\email[]{}
\affiliation{Department of Applied Mechanics, IIT Madras, 
Chennai 600~036}
\author{Sudhavathani Simon}
%\email[]{}
\affiliation{Department of Computer Science, Women's Christian College, 
Chennai 600~006}
\author{F. Gori and Massimo Santarsiero}
%\email[]{}
\affiliation{Dipartimento di Fisica, Universit\'{a} Roma Tre and 
CNISM, Vila della Vasca Navale~84, I-00146 Rome}
\author{Riccardo Borghi}
%\email[]{}
\affiliation{Dipartimento di Electronica Applicata, Universit\'{a} Roma 
Tre and CNISM, Vila della Vasca Navale~84, I-00146 Rome}
\author{N. Mukunda}
%\email[]{}
\affiliation{Centre for High Energy Physics, Indian Institute of 
Science, Bangalore 560~012}
\author{R. Simon}
%\email[]{}
\affiliation{The Institute of Mathematical Science, Tharamani, Chennai 
600~113}
%\maketitle

%\date{August 27, 2006} 

\begin{abstract}
The issue raised in this Letter is classical
 in the sense of being quite ancient: which subset of $4\times 4$ 
real matrices should be accepted as physical Mueller matrices in 
polarization 
optics? Non-quantum entanglement between the polarization 
and spatial degrees of freedom of an electromagnetic beam is shown to 
provide the  physical basis to resolve this 
issue in a definitive manner.
\end{abstract}
\maketitle

Entanglement is traditionally studied almost exclusively in the 
context of quantum 
systems. However,  this notion is basically kinematic,  
 and so  is bound to present itself  whenever and wherever  
the state space of interest is the  tensor product of two (or more) 
vector spaces. 
 Polarization optics of paraxial electromagnetic beams happens to 
have  
precisely 
this kind of a setting, and so one should expect entanglement 
 to play a  significant role in this situation. It turns out that  
 entanglement in this  non-quantum 
setup is not just a matter of academic curiosity: we shall show in 
this 
 paper that {\em consideration of this non-quantum entanglement 
 resolves a fundamental issue in classical 
polarization optics}. It will appear that this issue could not have been 
resolved without explicit consideration of entanglement. 
 We begin by outlining the structure of classical polarization 
optics\,\cite{MandelWolf,Brosseau,Simon82,Kim-Mandel-Wolf}

\noindent
{\bf The Mueller-Stokes Formalism}\,: 
Traditional Mueller-Stokes formalism applies to plane 
electromagnetic waves or, more generally, to elementary beams (see 
below). If the wave propagates 
 along the positive $z$-axis, 
 the components $E_1,\,E_2$ of the transverse electric field 
along the $x$ and $y$ directions can be arranged into a column vector   
\begin{eqnarray}
 {\mbox{\boldmath$E$}}
\equiv  
\left[\begin{array}{c} E_1\\ E_2\end{array}\right]\in
{\cal C} ^2
\end{eqnarray}
called the Jones vector, analogous to the state vector of a qubit.  
 [A  scalar factor of the form
$e^{i(kz-\omega t)}$ has been suppressed.]  
 While ${\mbox{\boldmath$E$}}^{\dagger}{\mbox{\boldmath$E$}}
 =|E_1|^2+|E_2|^2$ is (a
measure of) the intensity, the ratio $\gamma=E_1/E_2$ of the (complex)
components
specifies the state of polarization.

When ${\mbox{\boldmath$E$}}$  is not deterministic,  
 the state of polarization  is described by the
  {\em coherency or polarization matrix}
\begin{eqnarray}
 \Phi  \equiv \langle 
{\mbox{\boldmath$E$}}
{\mbox{\boldmath$E$}}^{\dagger}\rangle 
=\left[\begin{array}{cc}
\langle E_1E_1^*\rangle &
\langle E_1E_2^*\rangle\\
\langle E_2E_1^*\rangle&
\langle E_2E_2^*\rangle
\end{array}\right],
\end{eqnarray}
where $ \langle\,\cdots\,\rangle$ denotes ensemble average. 
The two defining properties of the  
coherency 
matrix are hermiticity, $\Phi^{\dagger}=\Phi$, 
and positivity, $\Phi\ge0$: every $2\times2$ 
matrix obeying these two conditions is a valid coherency matrix.

It is clear that the intensity corresponds to 
${\rm tr}\,\Phi$, and 
fully polarized (pure) states  describable by Jones 
vectors ${\mbox{\boldmath$E$}}$ 
correspond to ${\rm det}\,\Phi=0$.  
 Partially polarized or mixed states correspond to $\det\,\Phi>0\,.$ 
 Thus coherency matrices are analogous to the density operators of a 
qubit. 

Since $\Phi$ is hermitian, it can be conveniently described as 
{\em real} linear combination of the four hermitian matrices 
$\tau_0
= 1_{2\times 
2},\;\tau_1=\sigma_3,\;\tau_2=\sigma_1,\;\tau_3=\sigma_2$\, which are 
mutually orthogonal, ${\rm tr}\,\tau_a\tau_b = 2\delta_{ab}\,:$
\begin{eqnarray}
 \Phi=\frac{1}{2}\sum_{a =0}^{3}S_{a}\tau_{a} ~\Leftrightarrow ~
S_{a}={\rm tr}(\tau_{a}\Phi)\,. 
\end{eqnarray}
The coefficients $S_{a} $ define the components of the {\em 
Stokes 
vector} $S\in \mbox{\boldmath$R$} ^4$.  
 The intensity equals $S_0={\rm tr}\,\Phi$.

While hermiticity of $\Phi$  is equivalent to reality of the Stokes 
vector 
$S \in \mbox{\boldmath$R$}^4$, the positivity conditions ${\rm 
tr}\,\Phi>0,~{\rm det}\,\Phi\ge0$
read, respectively, 
$ S_0 >0,\;\,S_0^2-S_1^2-S_2^2-S_3^2\ge 0$.
Thus, {\em permissible polarization states correspond to the 
positive light
cone and its interior} (solid cone). Pure states live on the surface 
of this cone.

{\em Typical systems of interest in polarization optics are 
  spatially  homogeneous (in the transverse plane)}, in the sense 
that 
their action 
is 
independent 
of the coordinates $(x,\,y)$. If such a system is 
deterministic and acts
linearly on the field amplitude, it is described by a complex
$2\times 2$ numerical matrix $J$,  the Jones matrix of the system:
\begin{eqnarray}
 J:\;\;\;
&&{\mbox{\boldmath$E$}}
   \to {\mbox{\boldmath$E$}}'
=J{\mbox{\boldmath$E$}} \;\;\Leftrightarrow\nonumber\\
    && \Phi \equiv 
\langle {\mbox{\boldmath$E$}}
{\mbox{\boldmath$E$}}^{\dagger}\rangle 
\to \Phi' = \langle {\mbox{\boldmath$E$}}^{\,\prime}
{\mbox{\boldmath$E$}}^{\,\prime\,\dagger}\rangle 
=J\Phi J^{\dagger}.~~\;
\end{eqnarray}
Such Jones systems are analogous to hamiltonian evolutions of a 
qubit;  
since the intensity $S_0 = {\rm tr}\,\Phi$ need not be preserved, 
$J$ need not be unitary. 
 It is clear that Jones systems map pure states ($\det\,\Phi = 0$) into 
pure states.

We can go from a pair of indices, each running over $1$ and $2$, to 
 a single index running over $0$ to $3$ and vice versa. Thus, the 
elements of $\Phi$ can be written as an associated column vector 
$\tilde{\Phi}$ with 
$\tilde{\Phi}_0 = \Phi_{11}$,   
$\tilde{\Phi}_1 = \Phi_{12}$,   
$\tilde{\Phi}_2 = \Phi_{21}$, and    
$\tilde{\Phi}_3 = \Phi_{22}$. The one-to-one relationship (3) between 
$S$ and $\Phi$ may thus be written as the vector equation     
\begin{eqnarray}
\left[\begin{array}{c}S_0\\S_1\\S_2\\S_3\end{array}\right]=
\left[\begin{array}{cccc} 1&0&0&1\\ 1&0&0&-1\\ 0&1&1&0\\
    0&i&-i&0\end{array}\right] 
\left[\begin{array}{c} \tilde{\Phi}_0\\
    \tilde{\Phi}_1\\ \tilde{\Phi}_2\\
    \tilde{\Phi}_3\end{array}\right].
\end{eqnarray}
Thus, while   $\Phi = \langle {\mbox{\boldmath$E$}} 
{\mbox{\boldmath$E$}}^{\dagger}\rangle$ the associated  
 column vector $\tilde{\Phi} \equiv \langle {\mbox{\boldmath$E$}} 
\otimes {\mbox{\boldmath$E$}}^*\rangle$.
 The $4\times 4$ numerical matrix $A$ exhibited  above is 
essentially unitary: $A^{-1} = \frac{1}{2}A^{\dagger}$. 

Optical systems of interest can be more general than the ones 
described by Jones matrices. Such a general system is said to be  
non-deterministic, and acts directly on the 
Stokes vector rather than through the Jones vector. It is specified 
by a $4\times4$ real matrix called the Mueller matrix, transforming 
the Stokes vectors linearly:
\begin{eqnarray}
 M:\;S \to S'=MS.
\end{eqnarray}
Mueller matrix of a Jones system $J$ will be 
called Mueller-Jones matrix $M(J)$. 

Since $M$ produces a linear transformation on $S$, the {\em linear 
invertible}  relationship (3) or (5) between $S$ and $\Phi$ implies 
that 
$M$ will induce a linear transformation $H^{(M)}$ on $\Phi$, which 
we may write in the form\,:
\begin{eqnarray}
H^{(M)}:\;\; \Phi \to \Phi^{\,\prime},\;\;\Phi^{\,\prime}_{ij} = 
\sum _{k\ell}H^{(M)}_{\,ik,j\ell}\,\Phi_{k\ell}\,.
\end{eqnarray}
The fact that $\Phi^{\,\prime}$ needs to be  hermitian for all 
hermitian $\Phi$ demands that the map or super-operator 
$H^{(M)}$,  
viewed as a $4\times 4$ matrix with $ik$ (going over $0$ to $3$)  
labeling the rows and $j\ell$ labeling the rows, be  hermitian. 
 Let us define a new matrix $B^{(M)}$ by permuting the indices of   
 $H^{(M)}$:
\begin{eqnarray}
B^{(M)} _{\,ij,k\ell} =
H^{(M)} _{\,ik,j\ell}\,.
\end{eqnarray}
i.e.,  $B^{(M)}$ is obtained from  $H^{(M)}$ 
 by {\em simply interchanging} 
 $H^{(M)} _{02}$ with $H^{(M)}_{10}$, 
 $H^{(M)}_{03}$ with $H^{(M)}_{11}$, 
 $H^{(M)}_{22}$ with $H^{(M)}_{30}$, and  
 $H^{(M)}_{23}$ with $H^{(M)}_{31}$. 
In terms of $B^{(M)}$, Eq.\,(7) transcribes into the vector equation   
$\tilde{\Phi}^{\,\prime}= B^{(M)}\tilde{\Phi}$, and in view of (5)  
 we have the 
invertible relationship 
\begin{eqnarray}
B^{(M)} = A^{-1}M A\,,\;\;
 M = A B^{(M)}A^{-1}\,.
\end{eqnarray}
Thus, the correspondence between 
real matrices $M$ and  hermitian matrices 
$H^{(M)}$, established through (8) and (9), is indeed one-to-one. 
Elements of $H^{(M)}$ in terms of those of $M$  can be 
found in Eq.\,(8) of Ref.\,\cite{Simon82}.

If the system described by $M$ is a Jones system 
with Jones matrix $J$, it is 
clear from the transformation law 
  $\Phi \to \Phi' =J\Phi J^{\dagger}$ given in (4) that 
$B^{(M)} = J\otimes J^{\,*}$ and, consequently,  
$H^{(M)} =\tilde{J}\tilde{J}^{\dagger}$, where 
$\tilde{J}$ is the column vector associated with 
the $2\times 2$ matrix $J$. Thus we arrive at  
the following result of fundamental importance\,\cite{Simon82}.

\noindent
{\em Proposition}~1\,: A Mueller matrix $M$ represents 
a Jones system if and only if the associated hermitian matrix 
$H^{(M)}$ is a one-dimensional projection. 
If $H^{(M)}$ is such a projection  
$\tilde{J}\tilde{J}^{\dagger}$, 
then $M=M(J)$,  $J$ being  the  $2\times 2$  matrix 
associated with the column vector $\tilde{J}$.

As a consequence, which is  mathematically trivial but 
quite important for the issue on hand, we have\,\cite{Kim-Mandel-Wolf}  

\noindent
{\em Proposition}~2\,: A real matrix $M$ can be realized 
as a positive sum (ensemble) of Mueller-Jones matrices   
 if and only if the associated hermitian matrix 
$H^{(M)}$ is positive semidefinite. 
If $H^{(M)} = \sum_k \tilde{J}^{(k)}\tilde{J}^{(k)\,\dagger}$, 
 then $M = \sum_k M(J^{(k)})$
 where $M (J^{(k)}) = A(J^{(k)}\otimes J^{(k)\,*})A^{-1}$ is the 
Mueller-Jones matrix associated with $J^{(k)}$.   

With this brief outline, we are now ready to  
describe the fundamental issue being addressed in this Letter.

\noindent
{\bf The Issue}\,: The  Mueller-Stokes 
formalism  takes as state space 
$\Omega^{({\rm pol})}$ 
the collection of all Stokes vectors:
\begin{eqnarray} 
\Omega^{({\rm pol})} = \{\,S\in 
\mbox{\boldmath$R$}^{\,4}\,\,
|\,\,S_0>0,\;\;S^TGS\ge0\,\}\,,
\end{eqnarray} 
where $G={\rm diag}\,(1,\,-1,\,-1,\,-1)\,$. 
Thus given a $4\times 4$ real matrix $M$, in order that 
it qualifies to be a Mueller matrix one should demand that 
it maps the state space  $\Omega^{({\rm pol})}$ 
 into itself. Let us denote by 
${\cal M}$ 
the collection of all (real ) $M$ matrices which map 
$\Omega^{({\rm pol})}$ 
{\em into} itself. While the set of $M$ matrices which map 
$\Omega^{({\rm pol})}$ {\em onto} itself is clearly the 
six-parameter family 
$SO(3,1)\,\cup\,GSO(3,1)$, where $SO(3,1)$ is the 
 proper orthochronous 
 Lorentz group, ${\cal M}$ is much larger; it is a sixteen-parameter 
family. 

Let us denote by 
${\cal M}^{(+)}$
 the collection of $M$ matrices which can be 
realized as positive sum 
of Mueller-Jones systems $M(J)$. It is clear that 
${\cal M}^{(+)}$ is contained in ${\cal M}$. 
The structure of ${\cal M}^{(+)}$ is fairly simple:  
 it is clear from Proposition~1\,\cite{Simon82}   that 
 elements of ${\cal M}^{(+)}$ are in one-to-one correspondence 
with nonnegative $4\times 4$ matrices\,\cite{Kim-Mandel-Wolf}.   
 But the structure of ${\cal M}$ is considerably more involved. 
 Owing  to a sequence of 
developments\,\cite{Sanjay92,Givens93,Mee93,Sridhar94,Rao98},  
which are surprisingly recent in relative terms, we now have 
a complete characterization of  ${\cal M}$. The basic tool has been 
orbits of $M$ under double-coseting by $SO(3,1)$\,\cite{Sridhar94}.

That elements of 
${\cal M}^{(+)}$ 
are  Mueller matrices 
is clear, for they are {\em realized} as convex sums of Jones systems. 
That $M$ matrices which fall outside ${\cal M}$ are not Mueller 
matrices is also clear, for they fail to map the state space 
 $\Omega^{({\rm pol})}$  into itself. 
{\em Thus the issue is really one about the 
grey domain `in between',  the complement of 
 ${\cal M}^{(+)}$ in  ${\cal M}$}:   
 are these $M$ matrices physical Mueller matrices? 

 By definition, members of this 
domain cannot be realized  as positive sums of Jones systems; 
 but they map $\Omega^{({\rm pol})}$  into itself. 
No one has come up with a scheme 
to realize them  physically. 
On the other hand there are Mueller matrices,   
extracted from actual experiments, which fall deep into this grey  
domain (see Ref.\,\cite{Sridhar94} for examples from Ref.\,\cite{Zyl87}).

There are two difficulties in simply dismissing these matrices 
 as unphysical: first, the experimenters 
{\em did not realize them} as convex sums of Jones systems, and so 
the fact that they fall outside 
${\cal M}^{(+)}$ cannot be enough reason 
to dismiss them; and secondly, within the Mueller-Stokes formalism 
{\em there seems to exist no additional qualification we can demand of 
a Mueller matrix, over and above the requirement that it should map  
 $\Omega^{({\rm pol})}$  into itself}. 

As a simple illustration of this grey region between  
${\cal M}^{(+)}$ and ${\cal M}$, let us consider  
 $M$ matrices of the diagonal form 
${\rm diag}\,(1,\,d_1,\,d_2,\,d_3)$. 
 It is clear that $M$ will map  
 $\Omega ^{({\rm pol})}$ into  $\Omega ^{({\rm pol})}$, and hence be 
in  ${\cal M}$,  
if and only if $d_k \,\le \,1,\;\, k=1,\,2,\,3$\,. 
 In the Euclidean space 
$\mbox{\boldmath$R$}^3$ 
spanned by the parameters 
$(d_1,\,d_2,\,d_3)$ 
 this corresponds to  the solid cube with vertices at 
$(\pm1,\,\pm1,\,\pm1)$.

The hermitian matrix $H^{(M)}$ associated with  
${\rm diag}\,(1,\,d_1,\,d_2,\,d_3)$ is, from (10), 
\begin{eqnarray}
H^{(M)} =\frac{1}{2} 
\left[\begin{array}{cccc} 
1+ d_1\, &
0&
0 &
d_2 + d_3\, \\
0 &
\,1  - d_1\, &
\,d_2 - d_3\,  &
0 \\
0 &
\,d_2 - d_3\,  &
\,1   - d_1\, &
0 \\
\,d_2 + d_3  &
0 &
0 &
\,1   + d_1 
\end{array}\right]\!. \;\;
\end{eqnarray}
Clearly,   $H^{(M)}\ge 0$ if and only if  
$-d_1-d_2-d_3 \le 1,\;\,-d_1+d_2+d_3  \le 1,\;\,
 d_1+d_2-d_3 \le 1$, and $d_1-d_2+d_3  \le 1$.
i.e., iff  
$(d_1,\,d_2,\,d_3)$ is in the solid  
 tetrahedron  
with vertices at $(1,\,1,\,1),\,(1,\,-1,\,-1),\,(-1,\,1,\,-1)$ and 
$(-1,\,-1,\,1)$.   

\noindent
{\em Proposition}~3\,: For  $M$ matrices of the 
 restricted form ${\rm diag}\,(1,\,d_1,\,d_2,\,d_3)$,  
$\,{\cal M}$ corresponds to the cubical region with vertices at  
$(d_1,\,d_2,\,d_3) 
= (\pm1,\,\pm1,\,\pm1)$, whereas 
${\cal M}^{(+)}$  corresponds to the inscribed solid tetrahedron  
with vertices at $(1,\,1,\,1),\,(1,\,-1,\,-1),\,(-1,\,1,\,-1)$ and 
$(-1,\,-1,\,1)$.

In this Letter we present a compelling  {\em physical  ground}   
which judges every $M$ matrix which is not an element of  
 ${\cal M}^{(+)}$  as unphysical. And this physical ground comes from 
consideration of entanglement between  the polarization 
and spatial degrees of freedom of an electromagnetic beam. 

\noindent
{\bf Non-quantum Entanglement}\,: 
 Let us  now go beyond plane waves and consider 
paraxial electromagnetic beams. 
The simplest  beam field  has, in a transverse
plane $z=$ constant described by coordinates $(x,y)\equiv 
{\mbox{\boldmath$\rho$}},$ the
form 
$\mbox{\boldmath$E$}({\mbox{\boldmath$\rho$}})=(E_1\hat{\mbox{\boldmath$x$}}
+E_2\hat{\mbox{\boldmath$y$}})\,\psi({\mbox{\boldmath$\rho$}})$, 
where 
$E_1,\,E_2$ are  complex constants, and the scalar-valued function 
$\psi({\mbox{\boldmath$\rho$}})$ may be assumed to be square-integrable over
 the  transverse plane: 
$\psi({\mbox{\boldmath$\rho$}}) \in L^2({\mbox{\boldmath$R$}}^2)$.   
$\hat{\mbox{\boldmath$x$}},\;\hat{\mbox{\boldmath$y$}}$
are unit vectors along the $x,\;y$ axes. 
 It is clear that 
 the polarization part              
$(E_1\hat{\mbox{\boldmath$x$}}+E_2\hat{\mbox{\boldmath$y$}})$ and the 
spatial dependence or 
modulation
part $\psi({\mbox{\boldmath$\rho$}})$ of such a beam are well 
separated, 
allowing one to focus attention  
on one aspect
at a time. When one is interested in only the modulation aspect, 
the part
$(E_1\hat{\mbox{\boldmath$x$}}+E_2\hat{\mbox{\boldmath$y$}})$ may be 
suppressed, thus 
leading to `scalar optics'.
 On the other hand, if the spatial
part $\psi({\mbox{\boldmath$\rho$}})$ is suppressed 
we are led to the traditional
polarization optics or Mueller-Stokes formalism for plane waves. 

Beams whose polarization and spatial modulation separate in the 
above manner will be called {\em elementary beams}. 
Suppose we superpose or add two such elementary beam fields
$(a\hat{\mbox{\boldmath$x$}}+b\hat{\mbox{\boldmath$y$}})\,
\psi({\mbox{\boldmath$\rho$}})$ and
$(c\hat{\mbox{\boldmath$x$}}+d\hat{\mbox{\boldmath$y$}})\,
\chi({\mbox{\boldmath$\rho$}})$. The 
result is not of the
elementary form $(e 
\hat{\mbox{\boldmath$x$}}+f\hat{\mbox{\boldmath$y$}})
\,\phi({\mbox{\boldmath$\rho$}})$,  for any
$e,~f,~\phi({\mbox{\boldmath$\rho$}}),$ unless either 
$(a,\,b)$ is proportional to $(c,\,d)$
 so that one gets committed to a {\em common} polarization, 
or $\psi({\mbox{\boldmath$\rho$}})$ and
$\chi({\mbox{\boldmath$\rho$}})$ are proportional so that one gets 
committed to a {\em fixed} 
spatial mode. Thus, {\em the set of elementary fields is not 
closed 
under superposition}.

 Since superposition principle is essential for optics, 
 we are  led to consider beam fields of 
the more general form
${\mbox{\boldmath$E$}}({\mbox{\boldmath$\rho$}})
=E_1({\mbox{\boldmath$\rho$}})\hat{\mbox{\boldmath$x$}}
+E_2({\mbox{\boldmath$\rho$}})\hat{\mbox{\boldmath$y$}}$,   
and consequently to pay attention 
to the implications of 
inseparability or entanglement of polarization and 
spatial variation.
  This  more general form is
 obviously closed under superposition. We may write 
${\mbox{\boldmath$E$}}({\mbox{\boldmath$\rho$}})$ as a 
{\em (generalised) Jones vector}
\begin{eqnarray}
 {\mbox{\boldmath$E$}}({\mbox{\boldmath$\rho$}})
=\left[\begin{array}{c}E_1({\mbox{\boldmath$\rho$}})\\
E_2({\mbox{\boldmath$\rho$}})
\end{array}\right],\;\;
~~E_1({\mbox{\boldmath$\rho$}}),\,\,
E_2({\mbox{\boldmath$\rho$}})
\in L^2({\mbox{\boldmath$R$}}^2)\,. 
\end{eqnarray}
The 
intensity at location 
${\mbox{\boldmath$\rho$}}$ corresponds to 
$|E_1({\mbox{\boldmath$\rho$}})|^2 + |E_2({\mbox{\boldmath$\rho$}})|^2$.
 This field is of the elementary or separable form iff
$E_1({\mbox{\boldmath$\rho$}})$ and $E_2({\mbox{\boldmath$\rho$}})$  
are linearly dependent (proportional to one another). Otherwise, 
polarization and spatial modulation are inseparably entangled. 

The point is that the set of possible beam fields in a transverse plane 
constitutes the {\em tensor product} space 
${\cal C}^2 \otimes L^2({\mbox{\boldmath$R$}}^2)$. 
 But the set of all elementary fields constitutes  
 just the {\em set product}  ${\cal C}^2 \times L^2({\mbox{\boldmath$R$}}^2)$  
of  ${\cal C}^2$  and $L^2({\mbox{\boldmath$R$}}^2)$, 
 and hence forms a {\em measure zero subset} of the tensor product 
${\cal C}^2 \otimes L^2({\mbox{\boldmath$R$}}^2)$. 
 In other words, in  beam fields represented by  {\em generic elements} 
of ${\cal C}^2 \otimes L^2({\mbox{\boldmath$R$}}^2)$  polarization and 
 spatial modulation  
 should be expected to be  entangled\,:  {\em Entanglement is not an 
exception; 
it is the rule in} ${\cal C}^2 \otimes L^2({\mbox{\boldmath$R$}}^2)$, 
 the space of pure states appropriate for electromagnetic beams. 

To handle fluctuating beams, we need the 
beam-coherence-polarization (BCP) matrix 
$\Phi({\mbox{\boldmath$\rho$}};\,{\mbox{\boldmath$\rho$}}') \equiv
\langle {\mbox{\boldmath$E$}}({\mbox{\boldmath$\rho$}})
{\mbox{\boldmath$E$}}
({\mbox{\boldmath$\rho$}}')^{\dagger}\rangle$\,\cite{Gori98}\,: 
\begin{eqnarray}
 \Phi({\mbox{\boldmath$\rho$}};{\mbox{\boldmath$\rho$}}')=
\left[\begin{array}{cc}
\langle E_1({\mbox{\boldmath$\rho$}})E_1({\mbox{\boldmath$\rho$}}')^*\rangle&
\langle E_1({\mbox{\boldmath$\rho$}})E_2({\mbox{\boldmath$\rho$}}')^*\rangle\\
\langle E_2({\mbox{\boldmath$\rho$}})E_1({\mbox{\boldmath$\rho$}}')^*\rangle &
\langle E_2({\mbox{\boldmath$\rho$}})E_2({\mbox{\boldmath$\rho$}}')^*\rangle
\end{array}
\right],
\end{eqnarray}
As the name suggests, the 
BCP matrix describes both the
coherence and polarization properties.
 It is a
generalization of the numerical coherency matrix of Eq.\,(2),
 now to the case of beam fields. 

It is clear from the very definition (13) of  BCP matrix that 
this matrix kernel, viewed as an operator from 
${\cal C}^2\otimes L^2({\mbox{\boldmath$R$}}^2) 
\to {\cal C}^2\otimes L^2({\mbox{\boldmath$R$}}^2)$, 
  is hermitian nonnegative:
\begin{eqnarray}
\Phi_{jk}({\mbox{\boldmath$\rho$}};{\mbox{\boldmath$\rho$}}')
=\Phi_{kj}({\mbox{\boldmath$\rho$}}';{\mbox{\boldmath$\rho$}})^*,
~~j,\,k&=&1,\,2;\nonumber\\
\int d^2{\mbox{\boldmath$\rho$}} \,d^2{\mbox{\boldmath$\rho$}}' 
{\mbox{\boldmath$E$}}({\mbox{\boldmath$\rho$}})^{\dagger}
{\Phi}({\mbox{\boldmath$\rho$}};{\mbox{\boldmath$\rho$}}')
{\mbox{\boldmath$E$}}({\mbox{\boldmath$\rho$}}')\ge 
0,&&
\end{eqnarray}
$\forall {\mbox{\boldmath$E$}}({\mbox{\boldmath$\rho$}}) \in {\cal C}^2 
\otimes L^2({\mbox{\boldmath$R$}}^2)$.   
These are the {\em defining
properties} of the BCP matrix: {\em every 
$2\times2$ matrix   
of two-point functions ${\Phi}_{jk}({\mbox{\boldmath$\rho$}};
{\mbox{\boldmath$\rho$}}')$ meeting just these
two conditions  is a valid BCP matrix of some beam of light}.

\noindent
{\bf Resolution of the Issue}\,:
In the BCP matrix (16), each of the four blocks 
${\Phi}_{jk}({\mbox{\boldmath$\rho$}};{\mbox{\boldmath$\rho$}}')
=
\langle E_j({\mbox{\boldmath$\rho$}})E_k({\mbox{\boldmath$\rho$}}')^*
\rangle$ is an (infinite-dimensional) operator 
$L^2({\mbox{\boldmath$R$}}^2) \to L^2({\mbox{\boldmath$R$}}^2)$.  For 
the issue on hand, however, it turns out to be  sufficient to 
restrict the spatial 
dependence to just two orthonormal spatial modes 
$\psi_1({\mbox{\boldmath$\rho$}}),\; 
\psi_2({\mbox{\boldmath$\rho$}})$. 
This amounts to considering in 
place of  $L^2({\mbox{\boldmath$R$}}^2)$ the two-dimensional space 
${\cal C}^2$, the linear span of $\psi_1({\mbox{\boldmath$\rho$}})$ 
and  
$\psi_2({\mbox{\boldmath$\rho$}})$, so that the BCP matrix  
${\Phi}({\mbox{\boldmath$\rho$}};{\mbox{\boldmath$\rho$}}')$
  is a positive operator mapping 
${\cal C}^2 \otimes {\cal C}^2 
 \to {\cal C}^2 \otimes {\cal C}^2$, the first  
${\cal C}^2$ being for the polarization degree of freedom and the second  
${\cal C}^2$ for the spatial degree of freedom. 

Choice of 
a product basis in 
${\cal C}^2  \otimes {\cal C}^2$  
 transcribes the BCP matrix into a 
hermitian nonnegative numerical $4\times 4$ matrix  
of four $2 \times 2$ blocks, with the polarization (Roman) indices 
labeling the blocks and the spatial  (Greek) indices labeling entries 
within each block\,\cite{Santarsiero07}.  
The following orthonormal product basis suggests itself naturally:
\begin{eqnarray}
\chi_{1\alpha}({\mbox{\boldmath$\rho$}}) 
= \left[\begin{array}{c}1\\0 \end{array}\right]
\psi_\alpha({\mbox{\boldmath$\rho$}}),\;\, 
\chi_{2\alpha}({\mbox{\boldmath$\rho$}})
= \left[\begin{array}{c}0\\1 \end{array}\right]
\psi_2({\mbox{\boldmath$\rho$}}), 
\end{eqnarray}
 for $\alpha=1,\,2$. 
Thus a Jones vector 
$E({\mbox{\boldmath$\rho$}})$ 
in ${\cal C}^2  \otimes {\cal C}^2$ necessarily has the form
 $\sum_{j=0}^1\sum_{\alpha=0}^1 C_{j\alpha}
 \chi_{j\alpha}({\mbox{\boldmath$\rho$}})$. It is of the separable
or elementary form if and only if 
$C_{11}/C_{12} = C_{21}/C_{22}$.  
It can be 
identified with a four-dimensional numerical column vector 
${\mbox{\boldmath$C$}}$ 
whose entries 
are the expansion coefficients $C_{j\alpha}$: 
${\mbox{\boldmath$C$}} = 
[\,C_{11},\,C_{12},\,C_{21},\,C_{22}\,]^T$.  
It follows that the corresponding 
 (pure state) BCP matrix 
 $\Phi({\mbox{\boldmath$\rho$}},{\mbox{\boldmath$\rho$}}') 
 = E({\mbox{\boldmath$\rho$}}) E({\mbox{\boldmath$\rho$}}')$ 
 can be identified with the (numerical)  projection matrix 
 $\widehat{\Phi} = {\mbox{\boldmath$C$}}{\mbox{\boldmath$C$}}^\dagger$. 
The $M$ matrix acts on  $\widehat{\Phi}$ through the associated 
hermitian matrix $H^{(M)}$\,:
\begin{eqnarray}
H^{(M)}:\;\; \widehat{\Phi} \to 
\widehat{\Phi}^{\,\prime},\;\;\widehat{\Phi}^{\,\prime}_{i\alpha,j\beta} = 
\sum _{k\ell}H^{(M)}_{\,ik,j\ell}\,\widehat{\Phi}_{k\alpha,\ell\beta}\,.
\end{eqnarray}
Note that the Greek indices are left unaffected, 
consistent with the (transverse) spatial homogeneity of $M$.
  
Now consider the special Jones vector 
$E({\mbox{\boldmath$\rho$}})
=\chi_{11}({\mbox{\boldmath$\rho$}})+ \chi_{22}({\mbox{\boldmath$\rho$}})$ 
 corresponding to 
${\mbox{\boldmath$C$}} = [\,1,\,0,\,0,\,1\,]^T$  and  
$ \widehat{\Phi} = {\mbox{\boldmath$C$}}{\mbox{\boldmath$C$}}^\dagger$. 
Since $C_{j\alpha} = \delta_{j\alpha}$, we have 
$\widehat{\Phi}_{j\alpha,k\beta} = \delta_{j\alpha} \delta_{k\beta}$. 
That is, the only nonzero elements of the $4\times 4$  matrix 
 $\widehat{\Phi}$ are  
$\widehat{\Phi}_{00}=\widehat{\Phi}_{03}
= \widehat{\Phi}_{30} =
\widehat{\Phi}_{33} =1$. 

 Given an $M$ matrix, let $\widehat{\Phi}^{\,\prime}$ be the result of the action 
of $M$ on $\widehat{\Phi}$. We have from (16)   
\begin{eqnarray} 
\widehat{\Phi}^{\,\prime}_{i\alpha,j\beta} = 
\sum _{k\ell}H^{(M)}_{\,ik,j\ell}\,\widehat{\Phi}_{k\alpha,\ell\beta} = 
H^{(M)}_{i\alpha,j\beta}\,.
\end{eqnarray}
This is our final result. 
Suppose $H^{(M)}$ is not positive then $\widehat{\Phi}^{\,\prime}$, the 
result of $M$ acting on the entangled Jones 
vector $E({\mbox{\boldmath$\rho$}})
=\chi_{11}({\mbox{\boldmath$\rho$}})+ 
\chi_{22}({\mbox{\boldmath$\rho$}})$,  
fails to be positive  and hence is unphysical, showing in turn that $M$ 
could not have been physical. Thus $H^{(M)}\ge 0$ is a necessary 
condition for $M$ to be a physical Mueller matrix. On the other hand, we 
have seen that if $H^{(M)} \ge 0$ then $M$ {\em can be realized} as a 
convex sum of Jones systems, showing that $H^{(M)}\ge 0$ is a sufficient 
condition for $M$ to be a Mueller matrix. We thus have 

\noindent
{\em Theorem}\,: The necessary and sufficient condition for $M$ 
to be a physical Mueller matrix is that the associated 
hermitian matrix 
$H^{(M)}\ge 0$. 
{\em Every physical Mueller matrix is a convex sum of Mueller-Jones 
matrices}. 

Thus $M$ matrices which map $\Omega^{({\rm pol})}$ 
into itself should be called {\em pre-Mueller matrices} rather than 
Mueller 
matrices. For, to be promoted to the status of  Mueller matrices they  
 needs to meet the 
 stronger condition  $H^{(M)}\ge 0$ arising from consideration 
of entanglement. In view of the rapidly growing current interest in an 
unified  approach to coherence and polarization in optics, it is hoped 
that 
our result  will stimulate further research into the kinematic role of 
entanglement in classical optics.


\begin{thebibliography}{} 
\bibitem{MandelWolf}
L. Mandel and E. Wolf, {\em Optical Coherence and Quantum Optics} 
(Cambridge University Press, 1995), Chap.\,6.

\bibitem{Brosseau}
C. Brosseau, {\em Fundamentals of Polarized Light: A Statistical 
Approach} (Wiley, N.Y., 1998).

\bibitem{Simon82}
R. Simon, 
Opt. Commun. {\bf 42}, 293 (1982).

\bibitem{Kim-Mandel-Wolf}
K. Kim, L. Mandel, and E. Wolf, 
J. Opt. Soc. Am. A {\bf 4}, 433 (1987). 

%\bibitem{Stokes}
%G.G. Stokes, Trans. Cambridge Philos. Soc. {\bf 9}, 399 (1852). 


\bibitem{Sanjay92}
M.S. Kumar and R. Simon, 
Opt. Commun. {\bf 88}, 464 (1992).


\bibitem{Givens93}
C.R. Givens and B. Kostinski, 
J. Mod. Opt. {\bf 40}, 471 (1993).

\bibitem{Mee93}
V.M. van der Mee, 
J. Math. Phys. {\bf 34}, 5072 (1993). 


\bibitem{Sridhar94}
R. Sridhar and R. Simon, 
 J. Mod. Opt. {\bf 41}, 1903 (1994). 



\bibitem{Rao98}
A.V. G.  Rao, K.S. Mallesh, and Sudha, 
 J. Mod. Opt. {\bf 45}, 955 (1998); {\bf 45} 989 (1998).

\bibitem{Zyl87}
J.J. van Zyl, C.H. Papas, and C. Elachi, 
IEEE Trans. Antennas Prop. {\bf AP-35}, 818 (1987). 



\bibitem{Gori98}
F. Gori, 
Opt. Lett. {\bf 23}, 241 (1998).

\bibitem{Santarsiero07}
M. Santarsiero, F. Gori, R. Borghi, and G. Gauttari, 
J. Opt. A: Pure Appl. Opt. {\bf 9}, 593 (2007). 

 \end{thebibliography}
\end{document}